\documentclass[12pt]{article} 
\usepackage{epsfig} \usepackage{amsfonts,amssymb,amsmath}
\usepackage{subfigure}
\usepackage{enumerate}
\newcommand{\al}{\ensuremath{\alpha}}
\newcommand{\ga}{\ensuremath{\gamma}}

\newcommand{\ka}{\ensuremath{\kappa}}
\newcommand{\la}{\ensuremath{\lambda}}
\newcommand{\La}{\ensuremath{\Lambda}}

\newcommand{\del}{\ensuremath{\partial}}
 \newcommand{\half}{\frac{1}{2}}
 
 \newcommand{\be}{\begin{equation}}
\newcommand{\ee}{\end{equation}} \newcommand{\m}{\mathcal}
\newcommand{\ba}{\begin{eqnarray}} \newcommand{\ea}{\end{eqnarray}}

\newcommand{\M}{\ensuremath{\mathcal{M}}}

\newcommand{\lab}[1]{\label{#1}}
\newcommand{\bib}[1]{\bibitem{#1}}

 \newcommand{\nc}{\newcommand}

\def\ie{\textit{i.e. }} \def\mn{{\mu\nu}} \def\tt{\textrm}

\nc{\aaa}[3]{{\ Astron.\ Astroph.\ }{{\bf #1},({#2}){#3}}}
\nc{\advp}[3]{{Adv.\ in\ Phys.\ }{{\bf #1}({#2}){#3}}}
\nc{\apl}[3]{{Appl. Phys. Lett. }{{\bf #1}{(#2)}{#3}}}
\nc{\apj}[3]{{Astrophys.\ J.\ }{{\bf #1} {(#2)} {#3}}}
\nc{\apjl}[3]{{Astrophys.\ J.\ Lett.\ }{{\bf #1} {(#2)} {#3}}}
\nc{\app}[3]{{\ Astrop.\ Phys..\ }{{\bf #1}, {(#2)} {#3}}}
\nc{\asp}[3]{{\ Astropart.\ Phys.\ }{{\bf #1} {(#2)} {#3}}}

\nc{\cmp}[3]{{  Comm.\ Math.\ Phys.\ }{{ \bf #1} {(#2)} {#3}}}
\nc{\cqg}[3]{{  Class.\ Quant.\ Grav.\ }{{\bf #1} {(#2)} {#3}}}
\nc{\epl}[3]{{  Europhys.\ Lett.\ }{{\bf #1} {(#2)} {#3}}}
\nc{\ijmp}[3]{{ Int.\ J.\ Mod.\ Phys.\ }{{\bf #1} {(#2)} {#3}}}
\nc{\ijtp}[3]{{ Int.\ J.\ Theor.\ Phys.\ }{{\bf #1} {(#2)} {#3}}}
\nc{\jhep}[3]{{ JHEP\ }{{\bf #1} {(#2)} {#3}}} \nc{\jmp}[3]{{  J.\
Math.\ Phys.\ }{{ \bf #1} {(#2)} {#3}}} \nc{\jpa}[3]{{  J.\ Phys.\ A\
}{{\bf #1} {(#2)} {#3}}} \nc{\jpc}[3]{{  J.\ Phys.\ C\ }{{\bf #1}
{(#2)} {#3}}} \nc{\jpg}[3]{{ J.~Phys.~G:~Nucl.~Part.~Phys.~}{{\bf #1}
{(#2)} {#3}}} \nc{\jap}[3]{{ J.\ Appl.\ Phys.\ }{{\bf #1} {(#2)}
{#3}}} \nc{\jpsj}[3]{{ J.\ Phys.\ Soc.\ Japan\ }{{\bf #1} {(#2)}
{#3}}} \nc{\lmp}[3]{{ Lett.\ Math.\ Phys.\ }{{\bf #1} {(#2)} {#3}}}
\nc{\lncim}[3]{{ Lett.\ Nuov.\ Cim.\ }{{\bf #1} {(#2)} {#3}}}
\nc{\mpl}[3]{{ Mod.\ Phys.\ Lett.\ }{{\bf #1} {(#2)} {#3}}}
\nc{\nat}[3]{{  Nature \ }{{\bf #1} {(#2)} {#3}}} \nc{\ncim}[3]{{
Nuov.\ Cim.\ }{{\bf #1} {(#2)} {#3}}} \nc{\npb}[3]{{ Nucl.\ Phys.\
}{{\bf B#1} {(#2)} {#3}}} \nc{\pr}[3]{{ Phys.\ Rev.\ }{{\bf #1} {(#2)}
{#3}}} \nc{\pra}[3]{{  Phys.\ Rev.\ }{{\bf A#1} {(#2)} {#3}}}
\nc{\prb}[3]{{  Phys.\ Rev.\ }{{\bf B#1} {(#2)} {#3}}} \nc{\prc}[3]{{
Phys.\ Rev.\ }{{\bf C#1} {(#2)} {#3}}} \nc{\prd}[3]{{  Phys.\ Rev.\
}{{\bf D#1} {(#2)} {#3}}} \nc{\prl}[3]{{ Phys.\ Rev.\ Lett.\ }{{\bf
#1} {(#2)} {#3}}} \nc{\plb}[3]{{  Phys.\ Lett.\ }{{\bf B#1} {(#2)}
{#3}}} \nc{\prep}[3]{{ Phys.\ Rep.\ }{{\bf #1} {(#2)} {#3}}}
\nc{\prsl}[3]{{ Proc.\ R.\ Soc.\ London\ }{{\bf #1} {(#2)} {#3}}}
\nc{\ptp}[3]{{  Prog.\ Theor.\ Phys.\ }{{\bf #1} {(#2)} {#3}}}
\nc{\ptps}[3]{{ Prog\ Theor.\ Phys.\ suppl.\ }{{\bf #1} {(#2)} {#3}}}
\nc{\physa}[3]{{ Physica\ A\ }{{\bf #1} {(#2)} {#3}}} \nc{\physb}[3]{{
Physica\ B\ }{{\bf #1} {(#2)} {#3}}} \nc{\phys}[3]{{ Physica\ }{{\bf
#1} {(#2)} {#3}}} \nc{\rmp}[3]{{ Rev.\ Mod.\ Phys.\ }{{\bf #1} {(#2)}
{#3}}} \nc{\rpp}[3]{{ Rep.\ Prog.\ Phys.\ }{{\bf #1} {(#2)} {#3}}}
\nc{\sjnp}[3]{{ Sov.\ J.\ Nucl.\ Phys.\ }{{\bf #1} {(#2)} {#3}}}
\nc{\jetp}[3]{{ JETP\ }{{\bf #1} {(#2)} {#3}}} \nc{\yf}[3]{{ Yad.\
Fiz.\ }{{\bf #1} {(#2)} {#3}}} \nc{\zetp}[3]{{ Zh.\ Eksp.\ Teor.\
Fiz.\ }{{\bf #1} {(#2)} {#3}}} \nc{\zp}[3]{{ Z.\ Phys.\ }{{\bf #1}
{(#2)} {#3}}} \nc{\zpc}[3]{{ Z.\ Phys.\ C\ }{{\bf #1} {(#2)} {#3}}}
\nc{\ibid}[3]{{\sl ibid.\ }{{\bf #1} {#2} {#3}}}

\begin{document}

\vskip 1cm

\begin{center}
{\Large \bf The Instability of Vacua in Gauss-Bonnet Gravity}
\end{center}
\vskip 1cm \renewcommand{\thefootnote}{\fnsymbol{footnote}}

\centerline{\bf
Christos Charmousis$^{a}$\footnote{Christos.Charmousis@th.u-psud.fr} and Antonio
Padilla$^{b}$\footnote{antonio.padilla@nottingham.ac.uk}}
\vskip .5cm

\centerline{$^a$ \it Laboratoire de Physique Theorique,}
\centerline{\it Universit\'e de Paris-Sud,  B\^atiment 210,
  91405 Orsay CEDEX, France}
\vskip .5cm
\centerline{$^b$ \it School of Physics and Astronomy,}
\centerline{\it University Park, University of Nottingham, NG7 2RD, UK} \vskip .5cm

\setcounter{footnote}{0} \renewcommand{\thefootnote}{\arabic{footnote}}


\begin{abstract}
Owing to the quadratic nature of the theory, Einstein-Gauss-Bonnet gravity
generically permits two distinct vacuum solutions. One solution (the
"Einstein" vacuum)  has a well defined limit as the Gauss-Bonnet
coupling goes to zero, whereas the other solution (the "stringy"
vacuum) does not.  There has been some debate regarding the stability
of these vacua, most recently from Deser \& Tekin who have argued that
the corresponding black hole solutions have positive mass and as such
both vacua are stable. Whilst the statement about the mass is
correct, we argue that the stringy vacuum is still perturbatively unstable. Simply put,  the stringy vacuum suffers from a ghost-like
instability that is not excited by the spherically symmetric black hole, but would be excited by any source likely to emit gravitational waves, such as a binary system. This result is
reliable except in the strongly coupled regime close to the Chern-Simons limit, when the two vacua are
almost degenerate. In this regime, we study instanton transitions
between branches  via bubble nucleation, and calculate the nucleation probability.  This demonstrates that there is  large mixing between
the vacua, so that neither of them can accurately describe the true
quantum vacuum. We also present a new gravitational instanton describing black hole pair production in de Sitter space on the Einstein branch, which is preferred to the usual Nariai instantons and is not present in pure Einstein gravity.
\end{abstract}

\newpage
\section{Introduction}
In four dimensions, the Einstein-Hilbert action, supplemented with an arbitrary cosmological constant, is the unique gravitational action giving rise to field equations that only depend on the metric and its first two derivatives~\cite{Lanczos}. This is no longer true in $D>4$ dimensions, when the Einstein-Hilbert action is merely a special case of the Lovelock action, satisfying the same property~\cite{Lovelock}. In $D$ dimensions, the Lovelock action is given by a linear combination of $[\frac{D-1}{2}]$ dimensionally extended Euler characteristics,
\be
\label{euler}
S_\tt{Lovelock}=\sum_{n=0}^{[\frac{D-1}{2}]} \alpha_n S_n, \qquad S_n=2^{-n}\int_\M d^D x \sqrt{-g}~ \delta^{[c_1 d_1\ldots c_{n} d_n]}_{[a_1 b_1\ldots a_{n}b_{n}]}R^{a_1 b_1}{}_{c_1 d_1} \ldots  R^{a_n b_n}{}_{c_n d_n}
\ee
For any $n$, $S_n$ is a $D$-form that corresponds to a dimensionally extended Euler characteristic. it is a topological invariant in $D = 2n$ (trivial for $D< 2n$), but becomes dynamical for $D>2n$, and in just the right way as to keep the field equations down to second order. This is because the higher order derivatives that inevitably appear under metric variation of (\ref{euler}) can all be written as total divergences. Lovelock gravity is dynamically different from Einstein gravity
for $D>4$, even though the theories  are indistinguishable in four dimensions or less (for a review see \cite{mousis}).

Clearly Lovelock theory is the non-trivial and unique classical generalisation
of GR in higher dimensions and this makes it
 interesting to study in its own right. Not surprisingly it has also some relevance to string theory. Higher order
curvature terms are in fact known to appear in the slope expansion of the heterotic string, generically introducing higher derivatives in the metric~\cite{string1}.
This  will inevitably lead to perturbative ghosts, which are  known to be absent in the full non perturbative version of string theory.
One can avoid this apparent inconsistency (to leading order) if the slope expansion takes the form of the Lovelock action~\cite{string2}.
Indeed, the second order curvature terms  are known to take the form of the Gauss-Bonnet combination~\cite{gross}. Because this corresponds to the non-trivial $n=2$ Lovelock term in (\ref{euler}), we are guaranteed that there are no higher derivatives in the effective string action, and therefore no ghosts to second order.

This has motivated a great deal of interest in Lovelock gravity, and Einstein-Gauss-Bonnet (EGB) gravity in particular (see, for example,~\cite{des, gbbhs, myers, yang, tekin, gbham, random}). This is the simplest of Lovelock's extensions of General Relativity, and is described by the following action
\be \lab{gbaction}
S_{GB}=\kappa \int_\M d^D x \sqrt{-g}(-2\Lambda+R+\alpha L_{GB} ), \qquad L_{GB}=R^2-4R_{ab}R^{ab}+R_{abcd}R^{abcd}
\ee
where $\kappa=1/16 \pi G$. The action (\ref{gbaction}) is the sum of the first three Lovelock forms, the first two giving rise to the well known Einstein-Hilbert term and cosmological constant, the third to the Gauss-Bonnet invariant, $\sqrt{-g}L_{GB}$. The latter term  is topological in four dimensions, and becomes  dynamical in $D>4$ dimensions. EGB gravity has been extensively studied in the braneworld context owing to a shared ``stringy''
and cosmological motivation~\cite{gbbranes1,idefix, gbbranes2,antonis} and as a modified ``scalar-tensor'' theory of gravity in 4 dimensions
(see for example \cite{luca}).

In this paper, we will restrict attention to  $D$ dimensional EGB gravity,
coupled to ordinary matter fields. The field equations are given by
\be
G_{ab}+\La g_{ab} +\al H_{ab}=\frac{1}{2\ka} T_{ab}
\ee
where $T_{ab}$ is the matter energy-momentum tensor, $G_{ab}$ is the Einstein tensor for the metric $g_{ab}$, and
\be
H_{ab}=2RR_{ab}-4R_{ac}R_b^c-4R_{acbd}R^{cd}+2R_{acde}R_b{}^{cde}-\half L_{GB} g_{ab}
\ee
is the second order Lovelock tensor~\cite{Lovelock}. Given the higher order nature of the theory (\ref{euler}), the vacuum field equations generically admit multiple solutions with maximally symmetry (up to the order $n$ in number). For EGB,  we therfore have up to two distinct maximally symmetric vacuum solutions, with two possible {\it effective} cosmological constants,
\be
\Lambda^\pm_\tt{eff}=\Lambda_{CS}\left(1 \pm \sqrt{1-\frac{2 \Lambda}{\Lambda_{CS}}}\right), \lab{Lameff}
\ee
where
\be
 \Lambda_{CS}=-\frac{1}{4\alpha} \frac{(D-1)(D-2)}{(D-3)(D-4)}
\ee
This is true even in the absence of a bare cosmological contant in the action.
For these vacua to be well defined, the bare cosmological constant must satisfy the bound, $\La/\La_{CS}\leq 1/2$. It is easy to check that $\La_\tt{eff}^+/\La_{CS}\geq 1 \geq \La_\tt{eff}^-/\La_{CS}$, with equality when the bound is saturated, $\La=\La_{CS}/2$. This is known as the  Chern-Simons limit (at least in odd dimensions) \cite{zanelli}, and corresponds to the case where the two roots actually coincide. Note that only the lower root has a smooth limit, $\La_\tt{eff}^- \to \La$, as $\alpha \to 0$, and as such is often refered to as the ``Einstein'' branch. In contrast, the upper root, $\La_\tt{eff}^+$ has no smooth limit as $\al \to 0$, and represents a distinct new feature of EGB  gravity that is completely absent in higher dimensional General Relativity. For this reason, this branch is often refered to as the ``stringy'' or ``Gauss-Bonnet'' branch.

Our main interest here lies in the physical relevance of these vacua and hence in their stability, in particular that of the Gauss-Bonnet vacuum. There has been much to-ing and fro-ing in the literature over the years regarding this issue~\cite{des, gbbhs, yang, tekin}. We will review some of the various arguments in the next section, before clarifying the situation, we hope,  once and for all. As a brief taster, note that whilst we  agree with Deser and Tekin's recent claims that the gravitational mass of spherically symmetric solutions always takes the same sign as the inertial mass of the source~\cite{tekin}, we strongly disagree with their conclusions regarding the classical stability of the vacua. We will explain in detail why the Gauss-Bonnet   vacuum is certainly {\it un}stable, at least perturbatively, and why this does not conflict with Deser and Tekin's calculations of  gravitational mass.  In short, the instability is due to a freely propagating tensor ghost that is not excited by the spherically symmetric solutions.

It turns out that the perturbative analysis of section \ref{tale} actually breaks down close to the Chern-Simons limit, due to strong coupling. At this point, we need to investigate Gauss-Bonnet vacuum stability using non-perturbative techniques. To this end, in section \ref{instantons}, we will focus on instanton solutions, corresponding to differing vacuum bubbles surrounded by  domain walls of ordinary matter. Happily, our description works best in the near Chern-Simons limit, in contrast to the analysis of section~\ref{tale}. By employing standard semi-classical techniques~\cite{semiclassical1, semiclassical2}, we will find that the mixing between branches is completely unsuppressed. Our results indicate that in the near Chern-Simons limit, an empty, homogeneous vacuum, on either branch, does {\it not} describe a stable quantum vacuum, as it becomes littered with bubbles of vacua from the other branch.

We conclude this paper with some discussion, presenting a novel everywhere smooth EGB gravitational instanton with a non-zero temperature. This solution exists only on the Einstein branch, but is absent in General Relativity. It can be  used to demonstrate the semi-classical instability of de Sitter space,   in addition to the standard Nariai instanton.

\section{Chronicles of a  stringy vacuum} \lab{tale}
Let us begin this section with a short account regarding the status of Gauss-Bonnet vacua and their physical relevance.
We begin, not with the vacuum itself, but with a spherically
symmetric source of inertial mass, $M$.
Now in Einstein gravity, we know that Birkhoff's theorem applies outside the source,
and that the corresponding solution is Schwarzschild  or Schwarzschild-(anti) de Sitter
\be \lab{schw}
ds^2=-V(r)dt^2+\frac{dr^2}{V(r)}+r^2d \Omega_{D-2}
\ee
where the potential is given by
\be  \lab{vgr}
V(r)=-\frac{2\Lambda r^2}{(D-1)(D-2)}+1-\frac{M}{(D-2)\Omega_{D-2} \ka r^{D-3}}
\ee
and $d \Omega_{D-2}$ is the metric on a unit $(D-2)$-sphere. The gravitational (ADM) mass, $E$ of this solution exactly matches the inertial mass of the source, \ie $E=M$~\cite{ADM}. As long as $M>0$, the singularity at $r=0$ is shielded by a horizon in accordance with cosmic censorship {\footnote{In $D=5$ there is critical bound for the mass beyond which the solution is singular but this does not affect the argument presented here}}. In  Gauss-Bonnet gravity, a generalised form of Birkhoff's theorem{\footnote{In the Chern-Simons limit, an extra class of degenerate solutions exist \cite{idefix}}} also holds~\cite{gbbranes1, idefix, gbbirkhoff}, and, as expected, there are two branches of Schwarzschild-like solutions, with potentials given by~\cite{des, gbbhs}
\be
V_\pm(r)=1-\frac{2\Lambda_{CS}r^2}{(D-1)(D-2)}\left(1 \pm \sqrt{1-\frac{2 \Lambda}{\Lambda_{CS}}-\frac{M(D-1)}{\La_{CS}\Omega_{D-2} \ka r^{D-1}}}\right), \lab{vpm}
\ee
Now, even for $M>0$, we immediately see that the cosmic censorship hypothesis will not always hold on the Gauss-Bonnet branch -- naked singularities can form even  for sources of positive inertial mass! The situation becomes even more bizarre when one considers the asymptotic behaviour of the potential (\ref{vpm})
\be
V_\pm(r) \sim -\frac{2\Lambda_\tt{eff}^\pm r^2}{(D-1)(D-2)}+1 \pm \frac{M}{(D-2)\Omega_{D-2} \ka r^{D-3}}+\ldots \lab{linV}
\ee
One is tempted to compare this with the potential (\ref{vgr}) in Einstein gravity,  and to {\it naively} conclude that the gravitational mass is given by
\be
E_\pm=\mp M \lab{Ewrong}
\ee
If this were correct, it would mean the solution on the Gauss-Bonnet branch had {\it negative} gravitational mass, even though the source had positive inertial mass! Again, if correct, this would certainly render the Gauss-Bonnet vacuum unstable, as it would not be a true local ground state of the theory. This weird behaviour was attributed to perturbations around the vacuum containing a ghost~\cite{des, gbbhs, yang}. To see the ghost, consider metric perturbations, $\delta g_{ab}=h_{ab}$ about the maximally symmetric vacuum solution, $\bar g_{ab}$, with effective cosmological constant, $\La_\tt{eff}$. Expanding the action (\ref{gbaction}) to quadratic order, we find that
\be
\delta_2 S=-\frac{1}{2} \ka  \left(1-\frac{\La_\tt{eff}}{\La_{CS}}\right) \int_\M d^D x \sqrt{-\bar g}~ h^{ab} \left(\delta_1 G_{ab}+\La_\tt{eff} h_{ab}\right)+\delta_2 S_m \lab{delS}
\ee
where $\delta_1 G_{ab}$ is the Einstein tensor expanded to linear order in $h_{ab}$, and $\delta_2 S_m$ is the quadratic contribution from the matter part of the action. Variation of this action gives the linearised field equations
\be
 \left(1-\frac{\La_\tt{eff}}{\La_{CS}}\right)\left(\delta_1 G_{ab}+\La_\tt{eff} h_{ab}\right)=\frac{1}{2\ka}\delta_1 T_{ab} \lab{lineom}
\ee
The Einstein branch is well behaved at this order, since $\La_\tt{eff}/\La_{CS} <1 $, and so the kinetic part of the gravitational action has the usual sign. Indeed, we see from the equations of motion (\ref{lineom}) that the linearised theory is the same as in Einstein gravity, with a positive effective Newton's constant. In contrast, on the Gauss-Bonnet branch, $\La_\tt{eff}/\La_{CS} >1$, so the kinetic term in the action has the opposite sign, indicating the presence of a ghost. We also see how the linearised theory corresponds to Einstein gravity with a {\it negative} effective Newton's constant. This corresponds to an {\it anti}-gravity theory and leads to gravitational {\it repulsion}, consistent with the expansion of the potential (\ref{linV}).

Because the kinetic term of a ghost has the wrong sign, we can think of it as carrying negative kinetic energy. As a result, it was originally claimed~\cite{des, gbbhs, yang} that the ghost was responsible for the negative gravitational energy stored in the gravitational field around a spherically symmetric source of positive mass, as indicated by equation (\ref{Ewrong}). However, this statement {\it cannot} be correct, at least perturbatively around the vacuum,  as we will now explain.

A ghost can only lower the gravitational energy of the gravitational field if it is freely propagating. Now, the perturbation about the Gauss-Bonnet vacuum can be split into tensors, vectors and scalars in the usual way. By direct comparison with Einstein gravity, which is equivalent up to a sign at linear order, we can conclude that only the {\it tensor} modes propagate freely on the vacuum. Put another way: there are no scalar or vector gravitational waves. So, for the energy of the gravitational field to be negative, the spherically symmetric solution (\ref{vpm}) must excite the negative energy gravitational waves coming from the vacuum tensor perturbations.  However, this is {\it not} the case.  The spherically symmetric solution only excites scalar modes. Since these do not propagate freely, by virtue of Birkhoff's theorem, they only carry the energy given to them by the source. The gravitational energy of the black hole can only depend on the overall sign of the gravitational Hamiltonian, and not on which branch you are on.

This statement is actually consistent with some of Deser and Tekin's more recent work~\cite{tekin}. By developing new techniques for calculating the
ADM and AD energy of quadratic gravity theories, they found that the correct energy expressions should actually be given by
\be
E_\pm=M, \lab{Eright}
\ee
rather than equation (\ref{Ewrong}). The gravitational energy is always positive for a source of positive inertial mass, whatever branch you happen to be on! These new expressions were later confirmed using the Gauss-Bonnet Hamiltonian~\cite{gbham}, and agree with the  reasoning given in the previous paragraph.

Deser and Tekin went on however, to claim that the positivity of the energy (\ref{Eright}) meant that the vacua on both branches were stable.  We do not agree with this claim. We have already explained how the spherically symmetric solutions only excite scalars. These do not propagate freely and  can only carry the energy given to them by the source itself. However, we cannot rule out the possibility that other, non-spherically symmetric, solutions exist that excite the freely propagating tensor modes, eg a binary system with a time dependent mass quadrupole moment. On the Gauss-Bonnet branch, these do indeed give rise to negative energy gravitational waves, rendering the Gauss-Bonnet background unstable.

This kind of instability can manifest itself already at a classical level. To see how this might occur consider, for example, a binary system of black holes, or indeed any other source capable of emitting gravitational waves in $D$ dimensions (see~\cite{oscar} for some recent discussion of gravitational waves and binary systems in higher dimensional GR). If our theory of gravity happens to contain freely propagating tensor ghosts, the energy of our source will increase as it emits waves carrying negative energy. For the case of the binary system, its orbital period will become quicker and quicker as more and more gravitational waves are emitted at an ever increasing rate!

Even in a true vacuum, with no matter excitations, a ghost will lead to a quantum instability if it is coupled to matter, as it will be spontaneously produced in the vacuum along with ordinary matter fields~\cite{menace}. Since the ghost carries negative energy, and the matter fields, positive energy, this can occur without any overall violation of energy conservation. For a Lorentz invariant theory with no UV cut-off, the rate of particle production diverges, and the vacuum is destroyed infinitely quickly (see~\cite{tanaka}).

Our conclusion, then, is that although the Einstein vacuum is stable, the Gauss-Bonnet vacuum is certainly unstable. This perturbative instability is due to a tensor ghost, which can propagate freely on the vacuum. It can be seen even at a classical level whenever there is a source of gravitational waves. In the absence of any excited sources, the vacuum is still quantum mechanically unstable as long as the ghost couples to ordinary matter fields. Note that the ghost is not excited by the spherically symmetric solutions given by (\ref{schw}) and (\ref{vpm}), so it comes as no surprise that Deser and Tekin ultimately found the gravitational energy of these solutions to be positive. Those solutions only excite scalars, which are not freely propagating, and therefore only carry energy given to them by the source.

In the interests of completeness, we note that all of these results rely on the fact that the overall gravitational coupling, $\ka=1/16\pi G>0$. If we reverse the sign of $\ka$ (without altering the matter lagrangian), the situation is reversed: the Einstein branch suffers from a ghost, whilst the GB branch is stable. Since the overall sign of the gravitational Hamiltonian is  also reversed, the black holes masses take the {\it opposite} sign to the inertial mass of the source, but again, do not depend on which branch you are on. This theory does not admit a sensible general relativistic limit as $\al \to 0$, as the Einstein-Hilbert piece always takes the wrong sign in the action, so we will not discuss it further.

We end this section with a note of caution. The  perturbative analysis we have presented here is not universally applicable.  In particular, we see that it will fall victim to strong coupling when the coefficient of the kinetic term in  (\ref{delS}) becomes very small, \ie as $\La_\tt{eff} \to \La_{CS}$. This corresponds to nearing the Chern-Simons limit, when higher order interaction terms become important, and one needs to go beyond perturbation theory in order to study vacuum stability. This will be the subject of the next section.

\section{Bubbling Gauss-Bonnet vacua} \lab{instantons}

In the last section, we saw how the Gauss-Bonnet vacuum suffered from a perturbative ghost instability. This result holds firm as long as we are not too close to the Chern-Simons limit, at which point perturbation theory breaks down due to strong coupling. In this section, we will study vacuum stability using non-perturbative phenomena.  We will be interested in instantons that represent tunnelling between vacua, through bubble nucleation. As we will see later, our description will actually work best close to the Chern-Simons limit, \ie precisely when the analysis of the previous section becomes invalid. The techniques for studying vacuum decay were developed by Coleman and his collaborators, first for QFT in Minkowski space~\cite{semiclassical1}, and later including the effect of four dimensional Einstein  gravity~\cite{semiclassical2}. We will now extend that analysis to $D$ dimensional Gauss-Bonnet gravity (see also the very recent study \cite{cai3}).

We begin by constructing the so-called ``bounce'' geometries, named after the corresponding entity in particle mechanics. Typically, these describe spherically symmetric bubbles of ``true'' vacuum expanding in the ``false'' vacuum. Occasionally, the reverse is possible, with bubbles of false vacuum forming inside the true vacuum with a small but non-zero probability~\cite{truefalse}. In the thin wall limit, which we will consider here, the two vacua are separated by a domain wall, composed of ordinary matter.

Let us denote the interior of the bubble by $\M_1$ and the exterior by $\M_2$. We will assume that for $i=1, 2$,  $\M_i$ is a  maximally symmetric solution to the vacuum field equations
\be
G_{ab}+\La_i g_{ab}+\al_i H_{ab}=0.
\ee
Note that at this stage we are {\it not} assuming that $\La_1=\La_2$ or $\al_1=\al_2$. We do this for the purpose of generality, since we might want think of Gauss-Bonnet gravity as an effective theory, with these quantities corresponding to the vevs of other fields. However,  our ultimate interest will lie in transitions between branches, and the stability of the vacua for a given set of vevs. So, although we will keep things general for the most part, we will occasionally assume that the bare parameters are indeed the same on either side of the wall.

The wall itself represents the common boundary of the two manifolds, and is denoted by $\Sigma=\partial \M_1=\partial \M_2$. We will assume that the wall  has tension $\sigma$, so that the junction conditions{\footnote{Junction conditions are uniquely determined in Lovelock theory due to the fact that equations remain second order and boundary terms to a manifold with boundary are uniquely defined as dimensionally extended Chern $D-1$ forms (see for example \cite{mousis}). The relevant boundary term for Gauss-Bonnet was given by Myers in \cite{myers}. Varying this boundary term gives the result of \cite{gbjunc}}} there are given by~\cite{gbjunc}
\be
\ka \Delta \left[K_\mn-K \ga_\mn+2\al \left(Q_\mn-\frac{1}{3} Q \ga_\mn \right)\right]=\frac{\sigma}{2}\gamma_\mn \lab{junc}
\ee
 where the jump operator is $\Delta X=X_2-X_1$ and  the averaging operator is $\langle X\rangle=(X_1+X_2)/2$. The extrinsic curvature of the wall in $\M_i$ is given by the Lie derivative  of the induced metric $\ga_\mn$, with respect to the unit normal, $n^a$,  to the wall pointing out of  $\M_1$  and into $\M_2$,
\be
K^{(i)}_\mn=\frac{1}{2} \m{L}_n \ga_\mn.
\ee
If the Riemann tensor on the wall is given by $\m{R}_{\mu\nu\al\beta}$, then, suppressing the label $(i)$,
\ba
Q_{\mu\nu} &=& 2K K_{\mu\lambda}K^\lambda{}_\nu + K_{\lambda\rho}K^{\lambda\rho} K_{\mu\nu}
- 2K_{\mu\lambda}K^{\lambda\rho}K_{\nu\rho} - K^2 K_{\mu\nu} \nonumber\\
&& +2K \m{R}_{\mu\nu}+\m{R} K_{\mu\nu} -2K^{\lambda\rho}\m{R}_{\lambda\mu\rho\nu}-4 \m{R}_{(\mu}^{\; \lambda} K_{\nu) \lambda}
\ea
Note that we have used Latin indices to describe the bulk coordinates, and greek indices to describe the wall coordinates. In particular,  it is convenient to introduce Gaussian-Normal coordinates $x^a=(\xi, x^\mu)$ relative to the wall, which is fixed at $\xi=\xi_0$. The bulk metric is then given by
\be
ds^2=g_{ab}dx^a dx^b=d\xi^2+\rho(\xi)^2\left[-d\tau^2+\cosh^2\tau d \Omega_{D-2}\right] \lab{bulkg}
\ee
Note that the expanding bubble wall corresponds to a de Sitter hyperboloid of constant radius, $\rho(\xi_0)$,  embedded in the bulk geometry. If the effective cosmological constant in $\M_i$ is written as $\La_\tt{eff}^{(i)}=-\frac{1}{2}(D-1)(D-2) k_i^2$, then the function $\rho(\xi)$ is given by
\be
\rho(\xi)=\frac{1}{k_1}\sinh(k_1\xi), \qquad 0\leq \xi\leq \xi_0 \lab{rho1}
\ee
in $\M_1$ and
\be
\rho(\xi)=\frac{1}{k_2}\sinh[k_2(\xi-\beta)], \qquad \xi_0 \leq \xi \leq \xi_\tt{max} \lab{rho2}
\ee
in $\M_2$, where
\be
\xi_\tt{max}=\begin{cases}\infty & \tt{for $k_2^2 \geq 0$} \\
\beta +\pi/|k_2| & \tt{for $k_2^2<0$}\end{cases} \lab{ximax}
\ee
Note that these formulas are valid, even when $k_i^2 \leq 0$, as long as  the appropriate limit or analytic continuation is understood. The boundary conditions at the wall are given by continuity of the induced metric
\be \lab{cont}
\Delta\left[ \frac{\sinh k\la}{k}\right]=0, \qquad \la_1=\xi_0, ~~\la_2=\xi_0-\beta
\ee
and the junction conditions (\ref{junc}). The latter gives the following expression for the wall tension
\be
\sigma=-2(D-2)\ka \Delta \left[\frac{\rho'}{\rho}\left\{1-\frac{2}{3}(D-3)(D-4) \al \left[ \left(\frac{\rho'}{\rho}\right)^2-\frac{3}{\rho^2}\right]\right\}\right]_{\xi=\xi_0} \lab{sigma}
\ee
This expression only really makes sense  if the thin wall approximation is a good one.  For this to be so, we require~\cite{semiclassical2}
\be \lab{thin}
\frac{{\Big |}\Delta \rho'(\xi_0){\Big |}}{\rho(\xi_0)} \ll \frac{1}{\rho(\xi_0)} ~\implies~{\Big|}\Delta [\cosh k \la]{\Big |}\ll 1~\implies{\Big  |}|k_2|\la_2-|k_1| \la_1{\Big |}\ll 1
\ee
Furthermore, the wall is assumed to be made up of ordinary fields, and therefore its energy momentum tensor, $T_{ab}=-\delta (\xi-\xi_0)\sigma \ga_\mn \delta^\mu_a \delta^\nu_b$, must satisfy the weak energy condition. This means that we require $\sigma \geq 0$ for viable bounce geometries. As a check, consider the case where the vacuum is Einstein on both sides of the bubble wall, and take the GR limit, $\al_1=\alpha_2=0$, so that $\sigma=-2(D-2)\ka  \Delta \left[ \cosh k\la\right]/\rho(\xi_0)$.  Assuming the thin wall approximation (\ref{thin}), we see that $\sigma \geq 0$, if, and only if, we have one of the following
\begin{enumerate}[(i)]
\item $\La_1 \leq \La_2 \leq 0$ \lab{ads}
\item $\La_1 \leq 0 \leq \La_2$ \lab{mix}
\item $0\leq \La_1 \leq \La_2, ~|k_1|\la_1 \leq |k_2| \la_2 \leq \frac{\pi}{2}$ \lab{ds1}
\item $0\leq \La_2 \leq \La_1, ~|k_2|\la_2 \geq |k_1| \la_1 \geq \frac{\pi}{2}$\lab{ds2}
\end{enumerate}
Cases (\ref{ads}) to (\ref{ds1}) are consistent with~\cite{semiclassical2}, and describe tunnelling  between  a false vacuum of large $\La$, to a true vacuum of smaller $\La$.  If both vacua are de Sitter, we see from case (\ref{ds2}) that tunnelling can occur in the opposite direction,  from true to false, but with a very low nucleation rate (see, for example~\cite{truefalse}).

We are primarily interested in transitions between branches in Gauss-Bonnet gravity. With this in mind, let us consider the case where the bare parameters are the same inside and outside of the bubble (\ie $\Delta \La=\Delta \al=0$). It then follows that
\be
\sigma=-\frac{2}{3}(D-2)(D-3)(D-4) \ka \al \left(\frac{\Delta \left[\cosh k\la \right]}{\rho(\xi_0)}\right)^3 \lab{sigma2}
\ee
The thin wall approximation (\ref{thin}) has extra significance here, since it corresponds to being close to the Chern-Simons limit. This is precisely when the perturbative analysis of the previous section was in danger of breaking down due to strong coupling effects. Because they have opposite regimes of validity, we see explicitly how the two  sections compliment one another perfectly.  On the one hand we have a perturbative analysis that works best far away from the Chern-Simons limit, whereas on the other hand we have a non-perturbative analysis that works best close to that limit.

 Given the expression (\ref{sigma2}), and assuming the thin wall limit approximation (\ref{thin}), we see that $\sigma \geq 0$, if, and only if, one of the following holds
\begin{enumerate}[(a)]
\item $\al>0, ~\La^\tt{eff}_1 \leq \La_2^\tt{eff} \leq 0$ \lab{effads+}
\item $\al>0, ~\La_1^\tt{eff} \leq 0 \leq \La_2^\tt{eff}$ \lab{effmix+}
\item $\al<0, ~0\leq \La_1^\tt{eff} \leq \La_2^\tt{eff}, ~|k_1|\la_1 \geq |k_2| \la_2  \geq   \frac{\pi}{2}$ \lab{effds1-}
\item $\al<0, ~\La_2^\tt{eff} \leq 0 \leq \La_1^\tt{eff}$ \lab{effmix-}
\item $\al<0, ~0\leq \La_2^\tt{eff} \leq \La_1^\tt{eff}, ~ |k_2|\la_2 \leq |k_1| \la_1 \leq \frac{\pi}{2}$ \lab{effds2-}
\end{enumerate}
Cases (\ref{effads+}) to (\ref{effds1-}) describe tunnelling from high to low effective cosmological constant, consistent with decay of the false vacuum through the nucleation of a bubble of true vacuum. The remaining cases, (\ref{effmix-}) and (\ref{effds2-}), describe the reverse process, in which true vacuum decay occurs, and we have tunnelling to higher $\La_\tt{eff}$. These are analogous to case (\ref{ds2}) in the GR limit, although now we no longer require both vacua to be de Sitter.

Which is the true and false vacuum depends on the sign of $\al$.  For $\al>0$, the Gauss-Bonnet vacuum may be thought of as the true vacuum, as we always have $\La^+_\tt{eff} \leq \La_{CS} \leq \La_\tt{eff}^-$, whereas for $\al<0$, it is the Einstein vacuum that is the true vacuum, as  $\La^-_\tt{eff} \leq \La_{CS} \leq \La_\tt{eff}^+$. Note that while case (\ref{effds1-}) corresponds to the nucleation of a bubble of Einstein vacuum in Gauss-Bonnet, all other cases correspond to the reverse: bubbles of Gauss-Bonnet vacuum in Einstein.

Vacuum decay occurs when a region of  vacuum is replaced  with a region of another vacuum, such that there is a net energy loss inside the bubble. This energy is used to excite the fields that form the bubble wall, so that overall, energy is conserved.  We can illustrate this explicitly in the present case by calculating the energy of the bubble relative to the region of initial vacuum it replaced. This can be done using the Hamiltonian formulae derived in~\cite{gbham}. Note that in (\ref{bulkg}) we have expressed the bulk metric using cosmological coordinates along the direction of the wall. Since we want to calculate an energy, its convenient to transform to global de Sitter coordinates along the wall, so that the bulk metric is written as
\be
ds^2=d\xi^2+\rho(\xi)^2\left[-(1-r^2)dt^2+ \frac{dr^2}{1-r^2}+r^2 d \Omega_{D-3}\right] \lab{bulkg2}
\ee
Surfaces of constant $t$, $\Sigma_t$, provide a foliation of the bulk with respect to the timelike Killing vector $\del/\del t$. It follows that the lapse function $N=\rho\sqrt{1-r^2}$, and the shift vector vanishes. We also need to define the surface $S_t$, which is the intersection of the wall  with $\Sigma_t$, and has induced metric $q_{ij} dx^i dx^j=\rho^2(\xi_0)\left(\frac{dr^2}{1-r^2}+r^2 d \Omega_{D-3}\right)$. Now, from~\cite{gbham},  the energy stored inside the bubble is given by\footnote{The alert reader will notice that there is sign difference between equation (\ref{ebub}), and the corresponding energy formula (98) in~\cite{gbham}. In~\cite{gbham}, for any quantity $X$ in the test spacetime, with corresponding value $\bar X$ in the background, $\Delta X=X-\bar X$. In this paper, however, the bubble is given by $\M_1$, and so $X=X_1$ and $\bar X=X_2$, which means our definition of $\Delta$ differs by a sign, compared with~\cite{gbham}.}
\be
E_\tt{bubble}=\ka \int_{S_t} d^{D-2} x ~\Delta \m{L}^*_\tt{bdy}\lab{ebub}
\ee
where, for vanishing shift vector,
\be
 \m{L}^*_\tt{bdy}=\sqrt{q}N \left[2 \m{K} +12 \al \delta_i^{[l} \delta_j^m \delta_k^{n]}\left(\m{K}^i_l\left(\m{R}^{jk}{}_{mn}(q)-2 \m{H}^j_m\m{H}^k_n \right)-\frac{2}{3} \m{K}^i_l \m{K}^j_m \m{K}^k_n\right)\right]
\ee
Here $\m{R}_{ijkl}(q)$ is the Riemann curvature on $S_t$, and $\m{K}_{ij}$ and $\m{H}_{ij}$ are the extrinsic curvatures of $S_t$ in $\Sigma_t$ and the wall respectively. After a little algebra and integration, one finds that the energy inside the bubble is given by
\ba
E_\tt{bubble}&=&2\ka \Omega_{D-3}\rho(\xi_0)^{D-1} \Delta \left[\frac{\rho'}{\rho}\left\{1-\frac{2}{3}(D-3)(D-4) \al \left[ \left(\frac{\rho'}{\rho}\right)^2-\frac{3}{\rho^2}\right]\right\}\right]_{\xi=\xi_0} \\
&=&-\frac{\sigma}{D-2}\Omega_{D-3}\rho(\xi_0)^{D-1}
\ea
where we have used equation (\ref{sigma}). When the tension is positive, the energy of the bubble is negative relative to the region of vacuum it replaced. This applies in each of the five cases (\ref{effads+}) to (\ref{effds2-}), regardless of whether or not the bubble is one of true or false vacuum. It couldn't be any other way, since the energy stored in the wall must be compensated for by an energy deficit inside the bubble. In fact,
\be
E_\tt{wall}=\int_{S_t} d^{D-2}x \sqrt{q} N\sigma=-E_\tt{bubble}
\ee
and so the net energy is indeed zero.

We will now apply standard semi-classical techniques~\cite{semiclassical1, semiclassical2} in order to derive the probability of bubble nucleation  in the general  case. This is given by
\be
\m{P} \propto e^{-B/ \hbar} \lab{rate}
\ee
where  the instanton action, $B$, is given by the difference between the Euclidean bounce action and background action
\be
B=S_\tt{bounce}-S_\tt{background} \lab{B}
\ee
A general bounce is made up of the bubble interior, $\M_1$, the exterior, $\M_2$, and the bubble wall, $\Sigma$. In Lorentzian signature it is described by equations (\ref{bulkg}) to (\ref{rho2}). To go to Euclidean signature we Wick rotate the time coordinate, $\tau \to i\tau_E$, so that the Euclidean bounce geometry is given by
\be
ds^2=g_{ab}dx^adx^b=d\xi^2+\rho(\xi)^2d\Omega_{D-1} \lab{gE}
\ee
As usual, the wall, which was a de Sitter hyperboloid in Lorentzian signature, corresponds to a $(D-1)$-sphere in Euclidean signature. Taking care to include the relevant surface terms in the action~\cite{GH, myers}, we find that the Euclidean bounce action is given by
\ba
S_\tt{bounce}&=&-\ka \int_{\M_1 \bigcup \M_2}  \sqrt{g}~\left(R-2\La+\al L_\tt{GB}\right)+\int_\Sigma  \sqrt{\ga}\left[\ka\left(2\Delta K+\frac{4}{3} \Delta (\al Q)\right)+\sigma\right] \nonumber\\
&& \qquad \qquad \qquad +~( \tt{boundary terms at infinity}
)
\ea
To describe the background geometry, $\bar \M$, we simply extend the solution in $\M_2$ all the way to the centre of the space, so that the Euclidean background metric is given by
\be
d s^2=\bar g_{ab}dx^adx^b=d\xi^2+\bar \rho(\xi)^2 d\Omega_{D-1}
\ee
where
\be
\bar \rho(\xi)= \frac{1}{k_2}\sinh[k_2(\xi-\beta)], \qquad \beta \leq \xi \leq \xi_\tt{max}.
\ee
The background action is therefore given by
\be
S_\tt{background}=-\ka \int_{\bar \M}  \sqrt{\bar g}~\left(R-2\La+\al L_\tt{GB}\right)+~( \tt{boundary terms at infinity}
)
\ee
It is clear that when we calculate the instanton action, $B$, the boundary terms at infinity cancel one another. Plugging everything in, and making use of the expression (\ref{sigma}), we find that the instanton action is given by
\ba
B&=&\ka \Omega_{D-1} \Delta\left[\left(-2\La-D(D-1)k^2+D \ldots (D-3) \al k^4\right) \int_0^\la d\eta \left(\frac{\sinh k \eta }{k}\right)^{D-1}\right. \nonumber \\
&& \qquad \qquad \qquad \left.+2\rho^{D-2} \rho' \left\{1-2(D-2)(D-3)\al \left[\left(\frac{\rho'}{\rho}\right)^2-\frac{3}{\rho^2}\right]\right\}\right]_{\xi=\xi_0}
\ea
This expression is valid even when $\Delta \La \neq 0$, and/or $\Delta \al \neq 0$. Again, we use the GR limit as a check, taking both vacua to be Einstein and setting $\al_1=\al_2=0$ so that $B=2\ka(D-2)\Omega_{D-1} \Delta \left[\int_0^\la d \eta (k^{-1}\sinh k\eta)^{D-3}\right]$.  In the thin wall approximation (\ref{thin}),  we set $\La_i=\langle \La \rangle (1+\epsilon_i)$, where $|\epsilon_i|  \ll 1$, and then it is straightforward to check that $\langle \epsilon \rangle =0$. From (\ref{rho1}) and (\ref{rho2}), we see that $k_i \la_i =\sinh^{-1}k_i\rho(\xi_0)$, where $\rho(\xi_0)$ is fixed, and so
\be
B =-\frac{2}{D-1}\ka \Omega_{D-1} \rho(\xi_0)^{D}~f(\langle k \la \rangle)\Delta \Lambda+\m{O}(\epsilon^2)
\ee
where $f(z)=-\frac{1}{\sinh^2 z}\left(\frac{1}{\cosh z}-\frac{(D-2)}{\sinh z}\int^{z} _0dw~\left(\frac{\sinh w}{\sinh z}\right)^{D-3}\right)$. Now for $x, ~y \in \mathbb{R}$ we have $f(x) \geq  0$ for $x\geq 0$, whereas $f(iy) > 0$ for $y \in (0, \pi/2)$ and $f(iy) < 0$ for $y \in (\pi/2, \pi)$. As expected~\cite{semiclassical2}, $B>0$ for each case (\ref{ads}) to (\ref{ds2}), so the probability of  tunnelling between vacua is always exponentially suppressed. True vacuum decay, given by case (\ref{ds2}), is even further suppressed, as can be seen using the  identity $f(iy)+f(i(\pi-y))=-\frac{(D-2)}{sin^D y} \int_0^\pi dw \sin^{D-3} w$~\cite{truefalse}.

Returning to Gauss-Bonnet gravity,  we are mainly interested in transitions between branches,  and the tunnelling processes described by (\ref{effads+}) to (\ref{effds2-}).  Explicity setting $\Delta \La=\Delta \al=0$, we find that the instanton action simplifies, a little, to give
\ba
B&=&\ka \Omega_{D-1} \left\{\Delta\left[\left(-2\La-D(D-1)k^2+D \ldots (D-3) \al k^4\right) \int_0^\la d\eta \left(\frac{\sinh k \eta }{k}\right)^{D-1}\right]\right. \nonumber \\
&& \left. + 2\rho(\xi_0)^{D-1}\left[-2\left(\frac{D-1}{D-4}\right)\frac{\Delta \left[\cosh k\la\right]}{\rho(\xi_0)}+(D-2)(D-3)\al \left(\frac{\Delta \left[\cosh k\la\right]}{\rho(\xi_0)}\right)^3\right]\right\} \lab{Bmix}
\ea
As we have already stated, the thin wall approximation can only really be trusted close to the Chern-Simons limit, \ie when $\La_\tt{eff}^{(i)}=\La_{CS}(1 + \epsilon_i)$, where $|\epsilon_i| \ll 1$. For transitions between branches  the average effective cosmological constant, $\langle \La_\tt{eff} \rangle=\La_{CS}$, and so  $\langle \epsilon \rangle=0$. For cases (\ref{effads+}) to (\ref{effds2-}), it follows from (\ref{sigma2}) that $\sigma =\m{O}(\epsilon^3)>0$, and from (\ref{Bmix}) that $B=\m{O}(\epsilon^3)$.

Working to order $\epsilon$, we now see how a tensionless domain wall mediates transitions between branches in Gauss-Bonnet gravity, altering the effective cosmological constant.  At this order, the transition probability (\ref{rate}) is not suppressed, in contrast to similar transitions in General Relativity. This is the main result of this section, and can be interpreted as follows: close to the Chern-Simons limit, there is large mixing between the two distinct vacua in Gauss-Bonnet gravity. Neither the empty Einstein vacuum, nor the empty Gauss-Bonnet vacuum provide a good description of the stable quantum vacuum, as they will both  become littered with bubbles of the other.

It is interesting to note that a similar behaviour was seen in the DGP model~\cite{dgp}. In that case there is large mixing with empty five-dimensional space and one filled with so-called self-accelerating branes~\cite{dgppersp}. This pathology stemmed from the presence of perturbative ghosts on the self-accelerating branch~\cite{dgpghosts}, so given the results of the previous section, it is tempting to think that something similar is happening here. This may be so, but there are a number of reasons to tread carefully before making such a conclusion. Firstly, as we saw in the last section, the  perturbative ghost only haunts the Gauss-Bonnet branch, whereas here, tunnelling from {\it both} branches is unsuppressed.  Secondly, and most importantly, this section and the last have opposite regimes of validity, and in particular, one cannot really trust the existence of the ghost beyond the strong coupling scale, which is very low close to the Chern-Simons limit.

\section{Discussion}

Lovelock theory in $D>4$ spacetime dimensions differs dynamically from the naive extension of General Relativity in higher dimensions. Unlike GR and Chern-Simons gravity \cite{zanelli}, where the vacua are unique, Lovelock theory has multiple vacua for a given set of bare parameters in the action. In the simplest non-trivial extension of Lovelock theory, namely Einstein-Gauss-Bonnet (EGB) gravity, there are two such vacua: one  with a well defined limit as $\alpha \to 0$ (the Einstein vacuum), and one  without (the Gauss-Bonnet or ``stringy'' vacuum ). We have demonstrated that the Gauss-Bonnet vacuum is perturbatively unstable, despite the fact that the corresponding black hole excitations have positive mass,  as shown recently by Deser and Tekin~\cite{tekin}. The instability is due to a perturbative spin 2 ghost that is not excited in spherically symmetric vacua owing to an analogue of Birkhoff's theorem in Lovelock gravity~\cite{gbbranes1,idefix, gbbirkhoff}. However, since the ghost couples to matter, and other ordinary fields, it will inevitably lead to a quantum instability of the vacuum through ghost-nonghost pair production, without violating conservation of energy.

At the classical level, the pure vacuum itself is not unstable. A classical instability {\it can} appear if
a small amount of matter is put in by hand so that you are not quite in
vacuum. To see this, consider a matter source that is capable of
emitting gravitational waves such as a binary system of black holes with a time dependent mass quadrupole. Since the gravitons are ghost-like they carry
negative kinetic energy, and their emission leads to an increase in the
energy of the source. We soon end up with a spacetime containing a very high energy source surrounded by a turbulent sea of gravitational waves, and as such, this can no longer be regarded as a perturbation about the vacuum.  In this sense the introduction of a small matter source has rapidly destroyed the vacuum, destabilising the system already at the classical level.

It is natural to ask what is the end point of  the ghost instability? One might speculate that it leads to a transition between branches, which may certainly be possible at the quantum level.  We saw in section~\ref{instantons} how quantum transitions between branches {\it can} occur through bubble nucleation, although it was not clear whether or not this had anything to do with the ghost, or some other superselection mechanism. Another possibility is that the ghost mediates vacuum decay into spacetimes of less symmetry, as one might infer from the enhanced production of Nariai black holes in six dimensions~\cite{antonis}.  We will discuss this possibilty in more detail later on.  At the classical level, things are just as speculative. Classically, it is very difficult to see how a transition between branches could happen since it would require a change in the asymptotics. In this case, the end point, if it even exists,  will presumably depend on the nature of the source, and one may have to resort to complex numerical simulations to shed any more light on the issue.

All these statements regarding the (in)stability  of vacua at both the classical and quantum level are reliable as long as linearised perturbation theory works well on all but the shortest distance scales, \ie as long as $\La_\tt{eff}/\La_{CS} \ll 1$. However, as we approach the Chern-Simons limit in which the two vacua become degenerate ($\La_\tt{eff}/\La_{CS} \to  1$), the coupling of the graviton to other fields and to itself becomes stronger and stronger. This leads to a breakdown of linearised theory at larger and larger scales, pushing out the region of space in which we trust our perturbative analysis. Indeed, in the exact Chern-Simons limit, we see from (\ref{lineom}) that linearised theory breaks down altogether, on all scales. Close to this limit, we cannot make any robust conclusions regarding stability from a perturbative analysis alone. This limit has qualitative similarities to the partially massless limit for massive gravity in de Sitter~\cite{massiveds1, massiveds2}, aswell as  the chiral limit of topologically massive gravity\footnote{We thank Stanley Deser and Bayram Tekin for drawing our attention to this.} in AdS~\cite{tmg1, tmg2}, and the zero tension limit of DGP self-acceleration~\cite{dgp, dgppersp, dgpghosts}. In each case there is a ghost whose kinetic term seems to disappear~\cite{massiveds1, tmg1, dgpghosts} in the appropriate limit and a more careful analysis is required to identify any left over degrees of freedom. For massive gravity in de Sitter, the dangerous mode genuinely disappears~\cite{massiveds2}, whereas in the latter two cases the ghost mixes with another mode, such that one ghost-like degree of freedom remains~\cite{tmg2, dgpghosts}. We would need a direct  analysis beyond the scope of this paper to see what happens  in the exact Chern-Simons limit of EGB gravity.

Although linearised analysis breaks down close to this limit due to strong coupling, this is precisely the regime in which  we can reliably study instanton transitions between the two distinct vacua arising from the same theory (with the same bare parameters). The transitions are achieved through bubbles of one vacuum being nucleated within a sea of the other vacuum. The instanton analysis relies on the thin wall approximation, which works best near the Chern-Simons limit. This is because the two vacua are energetically very close to one another, and so only a small kink is required in the geometry of the bounce. In the thin wall/near Chern-Simons limit, we saw that the transition probability of bubble nucleation is not (linearly) suppressed unlike similar transitions in Einstein gravity \cite{semiclassical2}, and concluded that there is very strong mixing between the two distinct vacua of Gauss-Bonnet gravity.
For this reason, in the strong coupling regime, {\it neither} vacuum can accurately describe the true quantum vacuum state, since both will quickly become littered with bubbles of the other vacuum.

We have focussed on two pathologies associated with vacua in Einstein-Gauss-Bonnet gravity: a ghost-like instability in the perturbative regime on the GB branch, and unsuppressed mixing between vacua in the strong coupling regime. A third pathology was recently identified on the Einstein branch, through calculations of the viscosity-entropy bound \cite{visco} for gauge theories with a Gauss-Bonnet gravity dual. In the background of an EGB planar black hole, it was found that  graviton wave packets can propagate with superluminal speeds whenever $\frac{9}{25}\leq 8\alpha k^2 \leq 1$, where $1/k$ is the AdS length. This is not necessarily a sign of instability of the background, but one might expect some transition and novel physics to appear whenever $8\alpha k^2\sim \frac{9}{25}$. This scale does not show up in our calculations because we have considered perturbations of maximally symmetric backgrounds where the perturbation operator (\ref{lineom}) is simply a constant multiple of the Einstein perturbation operator. This is no longer the case when the background has a non-trivial Weyl tensor, as in the case of the planar black hole calculation of \cite{visco}.

We close off the section by complementing our instanton analysis of section~\ref{instantons} with some exact {\it gravitational} instantons present for positive effective cosmological constant. Some of these are present in General Relativity and correspond to the spontaneous creation of black hole pairs in the de Sitter vacuum, which are then accelerated away from one another by the background cosmological constant. In four dimensional GR, the  well known ``perdurance'' calculations~\cite{perry} demonstrate that de Sitter space is semi-classically unstable to the creation of Nariai black hole pairs, albeit at an exponentially suppressed rate (see~\cite{dias} for the extension to higher dimensional GR). In five dimensional EGB gravity, we find that the Nariai black hole can only exist on the Einstein branch, and only when the bare cosmological constant, $\La=6H^2>0$. The corresponding instanton has geometry $S^2 \times S^3$, and is given by the following metric
\be \lab{nariai}
ds^2=\left(\frac{1+8 \al H^2}{4H^2}\right) d \Omega_2+\frac{1}{2H^2} d\Omega_3
\ee
Recall that $\La/\La_{CS}\leq 1/2$ and so $1+8 \al H^2 \geq 0$. This means that the metric (\ref{nariai}) is well defined except in the Chern-Simon limit when the two-sphere shrinks to zero size. The background geometry is obtained by Wick rotating the Lorentzian de Sitter vacuum to euclidean signature. It is therefore given by a five-sphere of radius $H_\tt{eff}^{-1}$ where
\be
H^2_\tt{eff} =\frac{\La_\tt{eff}}{6}=\frac{\sqrt{1+8\al H^2}-1}{4\al}
\ee
is the effective curvature of the background. The instanton action (\ref{B}) is given by
\be
B_{Nariai}=4\pi \Omega_3 \ka\left(\frac{1+12 \al H_\tt{eff}^2}{H_\tt{eff}^3}-\frac{1+24 \al H^2}{\sqrt{2} H^3} \right)
\ee
which is always positive in the regime for which the solution (\ref{nariai}) is valid ($H^2>0, ~8 \al H^2 > -1$). The  creation rate of Nariai black hole pairs  is therefore exponentially suppressed, since ${\cal P} \propto e^{-B_{Nariai}/\hbar}$, as in General Relativity.

Interestingly, it turns out that there is an alternative channel for the creation of black hole pairs in five dimensional de Sitter, unique to EGB gravity. The process is described by a novel gravitational instanton solution that is,  again, only present  on the  Einstein branch, and has  metric
\be
\label{gravinst}
ds^2=V(r)d\tau^2+\frac{dr^2}{V(r)}+r^2 d\Omega_3^2
\ee
with
\be
V(r)=1+\frac{r^2}{4\alpha}\left(1-\sqrt{(1+8\alpha H^2)(1+\frac{16\alpha^2}{r^4})} \right)
\ee
The solution is only valid when  $\al>0$ and  $0<H^2 <1/8\al$, and is smooth everywhere in the range, $r_- \leq r \leq r_+$ where $r_\pm^2=\frac{1\pm\sqrt{1-64\alpha^2 H^4}}{2H^2}$ are the horizon positions, $V(r_\pm)=0$, for the corresponding Lorentzian black hole.  The smoothness of the solution (\ref{gravinst}) is guaranteed by the fact that $V'(r_+)=-V'(r_-)$, which means both horizons have the same temperature $T=|V'(r_\pm)|/4\pi$. Upon Wick rotating the Lorentzian black hole solution to Euclidean signature, this  ensures that the instanton solution (\ref{gravinst}) is free from conical singularities at both $r_-$ and $r_+$, provided we choose Euclidean time to have period $1/T$. The instanton (\ref{gravinst}) describes the sponteneous creation of the Lorentzian black holes in the de Sitter vacuum. The creation rate is again governed by the instanton action,
\be
B_{alternative}= 4 \pi \Omega_3 \ka \left[  \frac{1+12 \al H_\tt{eff}^2}{H_\tt{eff}^3}-(r_++r_-)^3       \right]
\ee
As before, this  is always positive within the regime of validity of the solution (\ref{gravinst}), so the corresponding black hole creation rate  is again exponentially suppressed. However, when both instantons solutions, (\ref{nariai}) and (\ref{gravinst}),  exist, it turns out that $B_{Nariai} \geq B_{alternative}$, with equality\footnote{After a suitable change of coordinates, it is possible to show that the solution (\ref{gravinst}) has a Nariai limit at the endpoint $H^2=1/8\al$, at which point the two horizons coincide} at the endpoint $H^2=1/8\al$. This means that the creation of these alternative black holes exceeds that of the Nariai black holes! In fact the creation rate of the alternative black holes  approaches order one as $\alpha \to 0$. This is because the alternative black hole, which has mass $M_{alternative} =6\al (1+8\al H^2) \Omega_3 \ka$, becomes arbitrarily small in the GR limit.

We wish to reiterate the fact that no such gravitational instantons exist on the Gauss-Bonnet branch in $D=5$ dimensions. However, in $D=6$, as was recently shown in \cite{antonis}, there exist Gauss-Bonnet branch Nariai instantons for which the instanton action is negative and therefore black hole production is exponentially enhanced! Naively,  this would seem to reflect the presence of the perturbative ghost discussed in section \ref{tale}. Of course, this ghost is present in $D=5$ aswell as $D=6$, so the absence of a five-dimensional Nariai instanton on the GB branch would seem to suggest otherwise. However, we can't rule out the possibility that there are other gravitational instantons whose action is also negative, leading to exponentially enhanced production rates.

To sum up we have demonstrated that the Gauss-Bonnet vacuum  is perturbatively unstable in EGB theory. We expect an analogous result to be true generically for all Lovelock theories. The result is robust as long as linearised  perturbation theory is valid. This is true beyond a certain length scale, which grows larger and larger as we approach the Chern-Simons limit. Close to that limit, when linearised perturbation theory has broken down almost completely, we found strong mixing between the two vacua, so that neither could accurately describe the true quantum vacuum.

\section*{Acknowledgements}
It is with great pleasure that we thank  Ed Copeland, Nemanja{ \tiny{d} }Kaloper,  Gustavo Niz, Antonis Papazoglou, Simon Ross, Paul Saffin, Lorenzo Sorbo for insightful comments and numerous discussions. We would particularly like to Stanley Deser and Bayram Tekin for their comments after reading through an earlier draft of this paper.  A{\tiny h}P would also like to thank Fernando Torres for making Spain Champions of Europe, and for making CC jealous!

\end{document}